\documentclass{article}
\usepackage[utf8x]{inputenc}

\usepackage{graphicx}

\usepackage{amssymb}   % Used for getting various symbols eg. gtrsim, lesssim etc.
\usepackage{amsmath}

\usepackage[compress]{cite}

\def\eq#1{{Eq.~(\ref{#1})}}

\def\cc{{cosmological\ constant}}
\def\LL{Lanczos-Lovelock }

\title{Exploring the Nature of Gravity}
\author{Thanu Padmanabhan\\
IUCAA, Pune University Campus,\\
 Ganeshkhind, Pune 411007, India\\
email: paddy@iucaa.in}

\date{ }

\begin{document}

\maketitle

\begin{abstract}
I clarify the differences between various approaches in the literature which attempt to link gravity and thermodynamics. I then
describe a new perspective based on the following features:  (1) As in the case of any other matter field, the gravitational field equations should also remain unchanged if a constant is added to the  Lagrangian; in other words, the field equations of gravity should remain invariant under the transformation $T^a_b \to T^a_b + \delta^a_b $(constant). (2) Each event of spacetime has a certain number ($f$) of microscopic degrees of freedom (`atoms of spacetime'). This quantity $f$ is  proportional to the area measure  of an equi-geodesic surface, centered at that event,  when the geodesic distance tends to zero. The  spacetime should have a zero-point length in order for $f$ to remain finite.  (3) The dynamics is determined by extremizing the heat density at all events of the spacetime. The heat density is the sum of a part contributed by matter and a part contributed by the atoms of spacetime, with the latter being $L_P^{-4} f$. The implications of this approach are discussed. 
\end{abstract}

\tableofcontents

%Elaborates on the results presented in five recent talks:  Keynote address in (i) EmQM15, Vienna, 23-25 Oct, 2015; Plenaries in (ii) 35th Max Born Symposium - The Planck Scale II, Wroclaw, 7-12 Sept, 2015 and (iii) ICGC-2015, Mohali, 14-16 Dec 2015; Colloquia at (iv) IAP, Paris, 30 Oct 2015 and (v) TIFR, Mumbai, 5 Oct 2015

\section{Linking Gravity with Thermodynamics: Comparison of  different approaches}

The idea that gravitational field equations could be interpreted using (or derived from)  thermodynamic arguments has been explored by many people  from widely different perspectives. (See e.g.,\cite{A26,A27,membrane,A35,tp02,tp04,aseemtp,tg17, tpmpla,tg16,lee2, A19,A11,tg14,lee1,A21,A20,hm258}).
There is a  tendency in the literature to club together these --- very different --- attempts as essentially the same or, at least, as  very similar. Such a point of view is
technically  incorrect, and given this tendency, it is useful to clarify the differences between the various approaches, as regards their assumptions, physical motivation and the generality  of the results. I will begin with a series of comments aimed at this task:\footnote{There are also numerous other attempts which derive/interpret  the \textit{linearized} field equations --- rather than the \textit{exact} equations --- from thermodynamic considerations. In what follows, I am only concerned with attempts which derive/interpret the \textit{exact} field equations.}

\begin{itemize}
 \item[(1)] To begin with, one must sharply distinguish between (i) the attempts concerned with the \textit{derivation} of the field equations by thermodynamic arguments (like e.g., \cite{A35,aseemtp,tg17,tpmpla,tg16,lee2,A19,tg14,lee1}) and (ii) the attempts related to the \textit{interpretation} of the field equations in a thermodynamic language (like e.g., \cite{A26,A27,membrane,tp02,tp04,A11,h256,tpsurf,A21,A20,A22,hm258}).  
 The latter is as important as the former because the existence of a purely thermodynamic interpretation for the field equations is vital for the overall consistency of the programme.
It is rather self-defeating  to derive the 
field equations $G^a_b = 8\pi T^a_b$ from thermodynamic arguments and then interpret them in the usual geometrical language! If gravity is thermodynamic in nature, then the gravitational field equations must be expressible in  a \textit{thermodynamic} language. This  crucial feature has not been given due recognition  in the literature. Unless the final result has an interpretation in thermodynamic language, such a derivation of the field equations is conceptually rather incongruous. 
 
 As an example of what I mean by such an interpretation, let me recall  the following result. It can be shown \cite{A19}  that 
 the evolution of geometry can be interpreted in thermodynamic terms, as the heating and cooling of null surfaces, through the equation:
\begin{equation}
\int_\mathcal{V} \frac{d^3x}{8\pi L_P^2}\sqrt{h} u_a g^{ij} \pounds_\xi p^a_{ij} = \epsilon\frac{1}{2} k_B T_{\rm avg} ( N_{\rm sur} - N_{\rm bulk})
\label{evlnsnb}
\end{equation} 
where
\begin{equation}
 N_{\rm sur}\equiv\int_{\partial \mathcal{V}} \frac{\sqrt{\sigma}\, d^2 x}{L_P^2};\quad
N_{\rm bulk}\equiv \frac{|E|}{(1/2)k_BT_{\rm avg}}
\end{equation} 
 are the degrees of freedom in the surface $\partial \mathcal{V}$ and bulk $\mathcal{V}$ of a 3-dimensional region  and $T_{\rm avg}$ is the average Davies-Unruh temperature \cite{du1,du2}  of the boundary. 
The  $h_{ab}$ is the induced metric on the $t=$ constant surface, $p^{a}_{bc} \equiv -\Gamma^{a}_{bc}+\frac{1}{2}(\Gamma^d_{bd}\delta^{a}_{c}+\Gamma^d_{cd}\delta^{a}_{b})$, and $\xi^a=Nu^a$ is the proper-time evolution vector corresponding to observers moving with four-velocity $u_a  = - N \nabla_a t$. The factor $\epsilon=\pm1$ ensures the correct result for either sign of the Komar energy $E$.
The time evolution of the metric in a region (described by the left hand side of \eq{evlnsnb}),  can be interpreted \cite{A21} as the heating/cooling of the spacetime and arises because $N_{\rm sur}\neq N_{\rm bulk}$. In any \textit{static} spacetime \cite{tpsurf}, on the other hand, $\pounds_\xi(...)=0$, leading to ``holographic equipartition'': 
$N_{\rm sur}= N_{\rm bulk}$.  This result translates gravitational dynamics  into  the  thermal evolution of the spacetime.
 The validity of \eq{evlnsnb} for all observers (i.e., foliations) ensures the validity of Einstein's equations.

In fact,  \textit{no} thermodynamic derivation of the  field equations in the literature  actually obtains the 
\textit{tensorial} form of field equation $G^a_b  = 8\pi T^a_b$. What is \textit{always done} is to obtain an equation of the form $G^a_b v_a v^b = 8\pi T^a_b v_av^b$ (where $v^a$ is either a timelike or null vector) and postulate its validity for all $v^a$. So it is important to understand the physical meaning of such an equation, especially the left hand side, for a given class of $v^a$. This will be a recurrent theme which I will elaborate on later sections.
 
 \item[(2)] Many thermodynamic derivations of field equations available in the literature, work with the assumption that the entropy of a horizon is proportional to its area (e.g., \cite{A35,tg16,lee2,tg14,lee1}) and attempt to introduce thermodynamic arguments centered around it. \textit{It is  likely that  such derivations miss some essential physics.}
 The connection between gravity and thermodynamics, motivated historically from the laws of black hole mechanics \cite{A1,A4} and the membrane paradigm \cite{A26,A27,membrane}, transcends Einstein's theory. In a  more general class of theories, the (Wald) entropy \cite{A8} of the horizon is \textit{not} proportional to its area.  One should therefore distinguish approaches in this subject which are specially tuned to Einstein gravity (and uses the entropy-area proportionality) from a broader class of approaches (like e.g. \cite{aseemtp,tg17,A20,A22}) because the latter ones, being more general, probably capture the underlying physics better.
  The above criticism is also valid for approaches based on entanglement entropy  when it is assumed to be proportional to the horizon area. 
 
 \item[(3)] Another feature which distinguishes different approaches in the literature is whether the field equations are derived from a variational principle or from some other procedure. I am personally in favour of approaches which use a variational principle because  they could offer a better window into microscopic physics. What is more, the approaches which does \textit{not} use a variational principle are very limited in their scope. For example, it is virtually impossible to generalize such models beyond Einstein's theory. (In contrast, the very first approach which used a thermodynamic variational principle to derive the field equations \cite{aseemtp}, obtained the field equations for all  \LL\ models at one go.) Some of these approaches like, for example, those which use the Raychaudhuri equation also have non-trivial technical issues \cite{dawood1,tg}.
 
 Even amongst the approaches which use variational principles, we need to distinguish between (i) those which vary the geometry (viz., the metric in some form, sometimes in a rather disguised manner) and (ii) those which vary some auxiliary vector field, keeping the metric fixed. Many approaches involving holographic concepts and entanglement entropy do vary the geometry in some form; however, I prefer approaches which vary an auxiliary vector. (After all, if you are going to vary the metric/geometry in an extremum principle, why not just use the Einstein-Hilbert action and be done with it~?!)   An example \cite{sanvedtp} of an extremum principle which does \textit{not} vary the metric,  is given by the functional
 \begin{equation}
 Q_{\rm tot}
=\int \sqrt{\gamma}\, d^2x \, d\lambda\, \left(T^a_b \ell_a\ell^b +\left[2\eta \sigma_{ab}\sigma^{ab}+\zeta\theta^2\right]\right)
\label{Qtotyy}
\end{equation}
Here, $\sigma_{ab}$ and $\theta$ are the shear and expansion of a null congruence $\ell^a(x)$, 
 $\eta=1/16\pi L_P^2,\ \zeta=-1/16\pi L_P^2$ are the shear and bulk viscous coefficients of a null fluid \cite{A26,A27,membrane} and the integrand can be interpreted as the rate of generation of heat (`dissipation without dissipation'; see \cite{h214,stefano}) due to matter and gravity on a null surface. Varying $Q_{\rm tot}$ with respect to $\ell^a$ and demanding that the extremum should hold for all $\ell^a$ (i.e., for all null surfaces) will lead to Einstein's equations. (We will say more about this in Sec. \ref{sec:dwd}.) Such a variational principle --- and others of a similar genre which we will discuss later --- treat the geometry as  fixed  and does not vary the metric. 

\item[(4)] At a more fundamental level, the horizon entropy cannot be finite unless some kind of discreteness exists in the spacetime near Planck scales. This is  clear in the case of entanglement entropy, which is a manifestly divergent quantity (see e.g., \cite{soloreview,tpentangle}) and needs to be regularized by some ad-hoc cut-off; but it is implicit in all  approaches. So, unless we have a model which captures \textit{at least some of} the quantum gravitational effects on the spacetime, any derivation of the field equations using a finite value for entropy is, at best, incomplete. 
 
 \item[(5)] Finally, let me emphasize that \textit{gravity cannot be an entropic force}. This was ably demonstrated by 
  Matt Visser \cite{mv} by an argument which uses (essentially) elementary vector analysis. It is trivial to prove,  in the Newtonian limit, that a conservative force ${\bf f} = -\nabla\phi$ cannot, in general,  be expressed in the entropic form ${\bf f} = T \nabla S$ if $T$ is the Davies-Unruh temperature that depends on the magnitude of the acceleration $|\nabla \phi|$. The relation $-\nabla\phi = T \nabla S$ implies that the level surfaces of $\phi$ coincide with those of $S$, allowing us to introduce a function $S= S(\phi)$. This, in turn, implies that $T(dS/d\phi) = -1$ and hence the level surface of $\phi$ coincide with the level surfaces of $T$. But since $T$ depends only on  $|\nabla \phi|$, this requires the level surfaces of  $|\nabla \phi|$ to coincide with those of $\phi$.  This condition is, in general, impossible to satisfy and can happen only in situations of high symmetry (for example, spherical, cylindrical, planar etc.). It would be preferable if the phrase ``entropic gravity'' is \textit{not used as a rather generic term} to describe the different approaches in this subject, for the simple reason that gravity cannot be an entropic force.

\end{itemize}

To summarize, there exist many different attempts in the literature to link gravity and thermodynamics. All of these are \textit{not} equivalent --- either conceptually or technically --- and it is also likely that at least some of them are fundamentally flawed or incomplete. 

The approach I have been pursuing --- which I will describe here --- is marked by the following features: (1) Much of it works for a wide class of theories, more general than Einstein's gravity. In particular, the results hold for theories in which entropy is \textit{not} proportional to horizon area. (2) The field equations are derived from a thermodynamic  extremum principle in which the geometry is not varied but some other auxiliary vector field is varied. (3) The resulting field equations are interpreted in a thermodynamic language and not in a geometric language.  (4) The introduction of a zero-point length to the spacetime by quantum gravitational effects allows us to provide a microscopic basis for the variational principle which is used.

Here, I will concentrate on developing this perspective \textit{from first principles} in a  streamlined manner. Obviously this will require us to make some educated guesses  but I shall argue that these guesses are well-motivated and the results are quite rewarding. 
In particular, I will describe the following two aspects: 

\begin{itemize}
 \item I will demonstrate \cite{tpentropy} a deep connection between two aspects of gravity which are usually considered in the literature to be quite distinct. The first is the fact that gravity seems to be immune to the shift in the zero level of the energy, i.e, to the shift in the value of cosmological constant. Second is  the feature I mentioned above, viz.,  gravitational dynamics can be reinterpreted in a purely thermodynamic language. I will show how the first feature \textit{leads to} the second and, in fact, provides a  simple and natural motivation to consider the heat density of the null surfaces as a key physical entity.

\item Much of the previous work  treated the spacetime as analogous to a fluid and investigated its properties in the \textit{thermodynamic limit}. The next, deeper, level of description of a fluid will be the \textit{kinetic theory}  which recognizes the discreteness and quantifies it in terms of a distribution function for its molecules.  I will describe an attempt \cite{tpentropy} to do the same for the spacetime by introducing a distribution function for the atoms of spacetime (which will count the microscopic degrees of freedom) and relating it to the extremum principle which, in turn,  will lead to the field equations. 
\end{itemize}

\section{Is the spacetime metric a dynamical variable?}

The principle of equivalence, along with principle of general covariance, strongly suggest that gravity is the manifestation of a curved spacetime\footnote{I use the signature $(- + + +)$ and will set $\hbar=1, c=1$ so that $G=L_P^2$. Occasionally, I will also set $G=1$ when no confusion is likely to arise.}, described by a non-trivial metric $g_{ab}(x)$. The \textit{kinematics} of gravity, viz. how a given gravitational field affects matter, can then be determined by postulating the validity of special relativistic dynamics in all freely falling frames.
This will lead to the condition $\nabla_a T^a_b =0$ for the energy momentum tensor of matter, which encodes the influence of gravity on matter.

Unfortunately, we do not have any equally elegant guiding principle  to determine the \textit{dynamics} of gravity, viz. how matter determines the evolution of the spacetime metric. The dynamics is contained in  the gravitational field equation, which --- in Einstein's theory --- is assumed to be given by $G^a_b = 8\pi T^a_b$. (In a more general class of theories, like e.g, \LL\ models, the left hand side will be replaced by a more complicated second rank, symmetric, divergence-free tensor.) One can obtain this equation, as Einstein did, by (i) assuming  that the right hand side \textit{must} be $T^a_b$ and (ii) by constructing a second rank, symmetric, divergence-free tensor from the metric containing upto second derivatives. Alternatively, as Hilbert did, one can write down a suitable scalar Lagrangian and vary it with respect to the metric and obtain the field equations.

In either procedure, one tacitly assumes that the spacetime metric is a dynamical variable with a status similar to, say, that of the gauge potential $A_j$ in electromagnetism. This belief is based on the fact that Einstein's equation is a second order differential equation for the metric just as Maxwell's equation is a second order differential equation for  $A_j$. It is in the same spirit that we justify varying the metric in the Hilbert action (as analogous to varying $A_j$ in the electromagnetic action) to obtain   Einstein's equation. Further, once we have a classical action $A_H[g_{ab}]$ in which $g_{ab}$ is varied to get the field equations, it is tempting to think of a quantum theory, defined through a path integral over $\exp iA_H[g_{ab}]$ 
 (or in some other  equivalent manner)
with the metric playing the role of a quantum variable.

But,  given the fact that spacetime geometry is conceptually very different from an external field propagating in it, this assumption --- viz., that the  metric is a dynamical variable similar to other fields --- is indeed nontrivial. Further, if varying the metric in the Hilbert action is \textit{not} the appropriate way to obtain the classical theory, then one is forced to think afresh about all the quantum gravity programmes.
Interestingly enough, this textbook procedure --- of treating  the metric as a dynamical variable,  
accepted without a second thought ---  is by no means a unique way to obtain  Einstein's equation. In fact, it is probably \textit{not} the most natural or efficient procedure. One can come up with alternative approaches \textit{and physically motivated extremum principles,} leading to Einstein's equation, in which the metric is \textit{not} a dynamical variable. Let me describe one such approach.

The field equations we seek should be a relativistic generalization of Newton's law of gravity $\nabla^2\phi\propto\rho$. 
A natural 
  way of generalizing this law is   to begin by noticing that: 
(i) The energy density in the right hand side $\rho=T_{ab}u^au^b$ is foliation/observer dependent where $u^i$ is the four velocity of an observer. There is no way we can keep $u^i$ out of it. 
(ii) We know from the principle of equivalence that $g_{ab}$ plays the role of $\phi/c^2$. So a covariant, scalar generalization of the left hand side, $\nabla^2\phi$, could come from the curvature tensor --- which contains the second derivatives of the metric. Any such generalization  \textit{must depend on the four-velocity $u^i$ of the observer} since the right hand side does. 
(iii) It is perfectly acceptable for the left hand side \textit{not} to have second \textit{time} derivatives of the metric, in the rest frame of the observer, since they do not occur in $\nabla^2\phi$. 

To obtain a scalar analogous to $\nabla^2\phi$, having \textit{spatial} second derivatives, we first project the indices of $R_{abcd}$ to the space orthogonal to $u^i$, using the projection tensor $P^i_j=\delta^i_j+u^iu_j$, thereby obtaining the tensor
$\mathcal{R}_{ijkl}\equiv P^a_iP^b_jP^c_kP^d_l R_{abcd}$. The only scalar we can get from $\mathcal{R}_{ijkl}$ is $\mathcal{R}^{-2}\equiv\mathcal{R}_{ij}^{ij}$ where $\mathcal{R}$ can be thought of as the radius of curvature of the space.\footnote{This $\mathcal{R}_{ijkl}$ and $\mathcal{R}$ should \textit{not} to be confused with the curvature tensor $^3R_{ijkl}$ and the curvature scalar $^3R$ of the 3-space orthogonal to $u^i$.} The natural generalization of Newton's law $\nabla^2\phi\propto\rho$ is then given by $\mathcal{R}^{-2}\propto\rho=T_{ab}u^au^b$. Working out the left hand side (see e.g., p. 259 of \cite{key7}) and fixing the proportionality constant from the Newtonian limit, one finds that
 \begin{equation}
G_{ab}u^au^b=8\pi T_{ab}u^au^b .  
\label{Guu}                               
\end{equation} 
If this scalar equation should hold for all observers (general covariance) then we need $G_{ab}=8\pi T_{ab}$ 
which is the standard result. Demanding that
  $\mathcal{R}^{-2}=8\pi\rho$ holds for each observer, captures the geometric statement --- viz. that energy density curves space as viewed by any observer ---  in a nice manner and is indeed the most natural generalization of  Newton's law: $\nabla^2\phi\propto\rho$. 

So, in this approach, the fundamental equation determining the geometry is  \eq{Guu} --- which should hold for all normalized, timelike vectors $u^i$ at each event of spacetime --- rather than the standard equation $G_{ab}=8\pi T_{ab}$.
While the two formulations are algebraically equivalent, they are conceptually rather different.  In the conventional approach to derive $G_{ab}=8\pi T_{ab}$, we do not  invoke any special class of observers. Instead, we \textit{assume} that the right hand side of the field equation \textit{must} be $T^a_b$ and  look for a generally covariant,  divergence-free, second-rank tensor built from geometry to put on the left hand side. (Alternatively, we look for a scalar Lagrangian made from geometrical variables). But the source in Newtonian gravity is actually $
T_{ab}u^au^b$ which \textit{does} involve an extra four-velocity for its definition. If we  introduce observers with four-velocity $u^i$ --- and in the end demand that the equation should hold for all $u^i$ --- we obtain the same gravitational field equations by a different route. This approach to \textit{dynamics} brings it closer to the way we handled the \textit{kinematics} by introducing the freely falling observers.

The real importance of this approach stems from the fact that it allows us to construct  a different kind of 
extremum principles which will lead to the gravitational field equations, \textit{without treating the metric as a dynamical variable!} Since this approach introduces an extra vector field $u^j$ into the fray, one can consider an extremum principle in which we vary $u^i$ --- which makes physical sense in terms of changing the observer --- instead of the metric. It is now possible, for example, to obtain the field equations by varying $u^i$ in a variational principle with the Lagrangian $L_1 \propto (G^a_b - 8\pi T^a_b) u_au^b$ and demanding that the extremum must hold for all $u^i$. In fact, we can also use the Lagrangian $L_2 \propto (R^a_b - 8\pi T^a_b) u_au^b$. Varying $u^j$ in the resulting action, after imposing the constraint $u^2 =-1$ and demanding that the extremum should hold for all $u^i$, will lead to the equation $R^a_b - 8\pi T^a_b = \lambda(x) \delta^a_b$ where $\lambda(x)$ is the Lagrange multiplier. Using  the Bianchi identity and $\nabla_a T^a_b =0$, we will recover the field equations except for an undetermined cosmological constant \cite{h191}. Removing a total divergence from $R^a_b u_au^b$, we see that this is equivalent to a variational principle based on the functional\footnote{This expression is very similar to the structure seen in ADM Hamiltonian (missing only a ${}^3R$ term) but, of course, here we are varying the vector field $u^i$ and \textit{not} the metric $g_{ij}$.  In fact, one can add any functional of the metric to the Lagrangian and it would make no difference since the metric is not varied.}
\begin{equation}
 A[u^i]  = \int d^4 x \, \sqrt{-g}\, \left[(\nabla_iu^i)^2 - \nabla_ju^i\nabla_iu^j - 8\pi \rho\right]
\label{tracekaction}
\end{equation} 
Varying $u_i$ in $A[u^i]$ and demanding the extremum to hold for all $u^i$ will lead to \eq{Guu} except for an undetermined cosmological 
constant.\footnote{Usually, if you vary a quantity $q_A$ in an extremum principle, you get an evolution equation for $q_A$. Here we vary $u_i$ in \eq{tracekaction} but get the equation constraining $g_{ab}$! This comes about because, after varying $u_i$, we demand that the equation  hold for all $u_i$ to take care of all observers. (Recall that this is also done in \textit{all} attempts to derive field equations by thermodynamic arguments.) While conceptually different from the usual extremum principles, it is perfectly well-defined and makes physical sense.}  So, one can indeed obtain the classical field equations for gravity \textit{without varying the metric} in any action principle.

The existence of such alternative variational principles takes away the motivation to treat the metric as a dynamical variable either classically or quantum mechanically. 
If this alternative procedure --- or a variant of it --- is the correct interpretation of classical gravity, then the  Hilbert action has no meaning classically (and hence in a quantum mechanical path integral). In the classical theory, what ultimately matters is the field equation and the rest is just window dressing.
But the distinction between these two approaches is vital when we want to bring together the principles of quantum theory and gravity. If  the metric is not a dynamical variable in the classical theory --- and the correct classical variational principle involves varying some other auxiliary variable like $u^i$ rather than the metric --- it makes no sense  to quantise the metric. Such an attempt will, at best, be similar to  quantizing the velocity or density field of a material medium. While gravitons will emerge with the same conceptual status as, say, phonons, the attempt will not lead to a complete quantum description of the spacetime. So, the alternative paradigm suggests a completely different picture about the quantum nature of spacetime. 

This point of view, that the metric is not  a dynamical variable, receives independent support from the tantalizing relationship between gravitational dynamics and the  thermodynamics of null surfaces. 
As I mentioned earlier, one can provide  a purely thermodynamic  interpretation to \eq{Guu} quite easily but \textit{not} to $G^a_b = 8\pi T^a_b$. We need the extra vector field $u^j$ for this interpretation, just as we need it to define an energy density. We will see later that the situation is still better when we use a null vector in place of the timelike vector.

\section{A guiding principle for dynamics and its consequences}

One major problem with the conventional approach to gravitational dynamics is that we lack a good guiding principle to determine the field equations. My first task is to take care of this.
I will begin by postulating  a guiding principle \cite{A19,A11}, which turns out to be as powerful as the principle of equivalence, in obtaining the  gravitational dynamics.

Let us recall that the equations of motion for matter, derived from an action principle, remain invariant if we add a constant to the matter Lagrangian, \textit{i.e.}, under the change $L_{m} \to L_{m} + $ constant. This encodes the principle that the zero level of energy density does not affect the dynamics. Motivated by this fact, it seems reasonable to postulate that the gravitational field equations should not break this symmetry, which is already present in the matter sector. Since $T_{ab}$ will occur, in one form or another, as the source for gravity (as can be argued from the principle of equivalence and considerations of the Newtonian limit), we postulate that:

\begin{itemize}

 \item[$\blacktriangleright$] The extremum principle that determines the dynamics of spacetime must be invariant under the change $T^a_b \to T^a_b + $ (constant) $\delta^a_b$.

\end{itemize}

\noindent This principle leads to two useful results:

First, 
this principle  rules out the possibility of varying the metric tensor $g_{ab}$ in a covariant, local, action principle to obtain the field equations! It is easy to prove \cite{C7} that if (i) the action is obtained from a local, covariant Lagrangian integrated over a region of spacetime with the covariant measure $\sqrt{-g}\, d^4x$ and (ii) the field equations are obtained through unrestricted variation\footnote{The second condition rules out unimodular theories and their variants, in which we vary the metric keeping $\sqrt{-g}$ fixed; I do not think we have a sound physical motivation for this approach.} of the metric in the action, then the field equations \textit{cannot} remain invariant under $T^a_b \to T^a_b + $ (constant) $\delta^a_b$.  In fact, $L_{m} \to L_{m} + $ constant is no longer a symmetry transformation of the action if the metric is treated as the dynamical variable. Therefore, any variational principle we come up with cannot have $g_{ab}$ as the dynamical variable.

As I highlighted in the last section, this need not scare us; it is certainly possible to come up with variational principles leading to the field equations in which the metric is not a dynamical variable. This brings us to the second result: The most natural structure, built from $T^a_b$, which maintains the required invariance under $T^a_b \to T^a_b + $ (constant) $\delta^a_b$, is given by
\begin{equation}
 \mathcal{H}_m[\ell_a] \equiv T_{ab} \ell^a \ell^b
\label{Qtot}
\end{equation} 
where $\ell_a$ is a null vector.\footnote{We want to introduce a minimum number of extra variables. In  $d$-dimensional spacetime, the null vector with $(d-1)$ degrees of freedom is the minimum we need. In contrast, if we use, say, a combination $T^{ab}V_{ab}$ with a symmetric traceless tensor $V_{ab}$, in order to maintain the invariance under $T^a_b \to T^a_b + $ (constant) $\delta^a_b$, then we need to introduce $(1/2)d(d+1)-1$ degrees of freedom; in $d=4$, this introduces nine degrees of freedom, which is equivalent to introducing three null vectors rather than one.}
This is very similar to the right  hand side of \eq{Guu} except  that we are now using a \textit{null} vector $\ell_a$ rather than a \textit{timelike} vector $u^i$ to implement our guiding principle. Demanding that the equation 
\begin{equation}
 G_{ab}\ell^a\ell^b= R_{ab}\ell^a\ell^b=8\pi T_{ab}\ell^a\ell^b 
\label{Gnn}
\end{equation}
holds for all null vectors $\ell_a$ at each event  
will also lead to $G^a_{b}=8\pi T^a_{b}+\Lambda\delta^a_b$ (where $\Lambda$ is the undetermined \cc) and our guiding principle points towards such an approach. While the right hand side of \eq{Guu} ($T^a_b u_au^b$) changes under  $T^a_b \to T^a_b + $ (constant) $\delta^a_b$, the  right hand side of \eq{Gnn}   ($T^a_b \ell_a\ell^b$) remains invariant, which is what we want. But,  $\rho\equiv T_{ab}u^au^b$ has a clear physical meaning as the energy density  measured by an observer with velocity $u^i$, but the physical meaning of $\mathcal{H}_m[\ell_a] \equiv T_{ab} \ell^a \ell^b$ is not obvious. Our next task  is to clarify that.

In the case of an ideal fluid, with $T^a_b = (\rho+p) u^au_b + p\delta^a_b$, the combination $T^a_b \ell_a\ell^b$ is actually the \textit{heat density} $\rho +p = Ts$ where $T$ is the temperature and $s$ is the entropy density of the fluid. (The last equality follows from Gibbs-Duhem relation and we have chosen the null vector with $(\ell.u)^2=1$ for simplicity.)  The invariance of $T^a_b \ell_a \ell^b$  under  $T^a_b \to T^a_b + $ (constant) $\delta^a_b$  reflects the fact that the cosmological constant, with the equation of state $p+\rho =0$, has  zero heat density. Our guiding principle, as well as \eq{Gnn}, suggests that \textit{it is the heat density rather than the energy density which is the source of gravity. }

But  $T^a_bu^bu_a$ is the energy density for \textit{any} kind of $T^a_b$, not just for that of an ideal fluid. How do we interpret $T^a_b \ell_a\ell^b$  in a  general context when $T^a_b$ could describe any kind of source --- not necessarily a fluid --- for which concepts like temperature and  entropy do not exist intrinsically?  \textit{Remarkably enough, this can be done}. In any spacetime, around any event, there exists a class of observers (called  a local Rindler observers) who will interpret $T^a_b \ell_a\ell^b$ as the heat density contributed by the matter to a null surface which they perceive as a horizon. This \textit{motivates} us to introduce \cite{A35} the concept of local Rindler frame (LRF) and local Rindler observers which will allow us to provide a thermodynamic interpretation of $T^a_b \ell_a\ell^b$ for any $T^a_b$. This arises as follows:

\begin{figure}
 \begin{center}
  \includegraphics[scale=0.4]{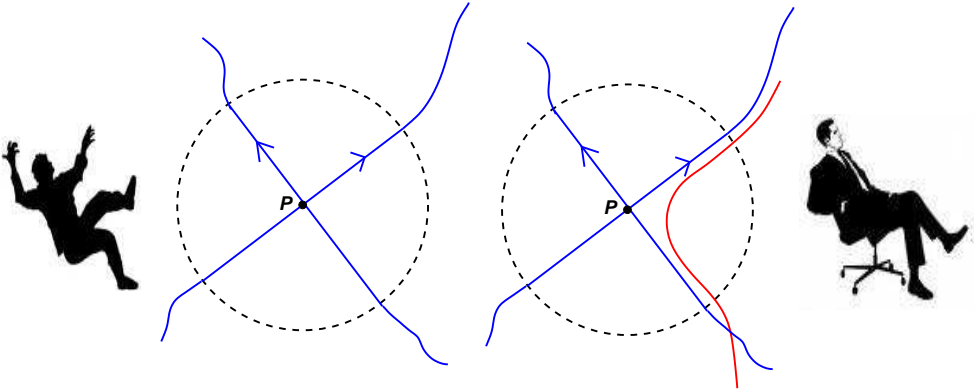}
 \end{center}
\caption{ (a) Left: A freely falling observer, with an associated local inertial frame, defined within the region marked by the black circle. The radius of this circle is decided by the curvature of the spacetime at $\mathcal{P}$.  
Light rays travelling at 45 degrees in the local inertial frame define the light cones at $\mathcal{P}$.
(b) Right: A local Rindler observer who is accelerating with respect to the inertial observer. For a sufficiently large acceleration, the trajectory of such an observer will be close to the light cones emanating from $\mathcal{P}$. The local Rindler observer will perceive the light cone as a local Rindler horizon and attribute to it a temperature given by \eq{rindlertemp}. In other words, the vacuum fluctuations of the local inertial frame will appear as thermal fluctuations in the local Rindler frame.}
\label{fig:daviesunruh}
\end{figure}

In a region around any event $\mathcal{P}$, we first introduce the freely falling frame (FFF) with coordinates $(T, \mathbf{X})$. Next, we boost from the FFF to a local Rindler frame (LRF) with coordinates $(t,\mathbf{x})$ constructed using some acceleration $a$, through the transformations: $X=\sqrt{2ax}\cosh (at), T=\sqrt{2ax}\sinh (at)$ when $X>|T|$ and similarly for other wedges. One of the null surfaces passing though $\mathcal{P}$, will get mapped to the $X=T$ surface in the FFF and will act as a patch of horizon to the $x=$ constant Rindler observers.
This construction leads to a beautiful result \cite{du1,du2} in quantum field theory. The local vacuum state, defined by the freely-falling observers around an event, will appear as a thermal state to the local Rindler observers with the temperature: 
\begin{equation}
 k_BT = \left(\frac{\hbar}{c}\right) \left(\frac{a}{2\pi}\right)
\label{rindlertemp}
\end{equation} 
where $a$ is the acceleration of the local Rindler observer, which can be related to other geometrical variables of the spacetime in different contexts [see Fig.~\ref{fig:daviesunruh}]. 
The existence of the  Davies--Unruh temperature tells us that around \textit{any} event, in \textit{any} spacetime, there exists a class of observers who will perceive the spacetime as hot.
 
Let us now consider  the flow of energy associated with the matter that crosses the null surface. Nothing unusual happens when this is viewed in the FFF by the locally inertial observer. But the local Rindler observer attributes a temperature $T$ to the horizon and views it as a hot surface. Such an observer will interpret the energy $\Delta E$, dumped on the horizon, by the matter that crosses the null surface,  as  energy  added to a \textit{hot} surface, thereby contributing a \textit{heat} content $\Delta Q=\Delta E$. (Recall that, as seen by the outside observer, matter actually takes an infinite amount of time to cross a \textit{black hole} horizon, thereby allowing for thermalization to take place. Similarly, a local Rindler observer will find that the matter takes a very long time to cross the horizon.) To compute $\Delta E$ in terms of $T^a_b$, note that the LRF provides us with an approximate Killing vector field $\xi ^{a}$, generating the Lorentz boosts, which coincides with a suitably defined null normal $\ell ^{a}$ at the null surface. The heat current arises from the  energy current $T_{ab}\xi ^{b}$ of matter and hence the total heat energy dumped on 
the null surface will be:
\begin{align}\label{Paper06_New_11}
 Q_{m}=\int \left(T_{ab}\xi ^{b}\right)d\Sigma ^{a}=\int T_{ab}\xi ^{b}\ell ^{a}\sqrt{\gamma}d^{2}x d\lambda
=\int T_{ab}\ell ^{b}\ell ^{a}\sqrt{\gamma}d^{2}x d\lambda
\end{align}
where we have used the result that $\xi ^{a} \to \ell ^{a}$ on the null surface. 
So we find that\footnote{Since null vectors have zero norm, there is an overall scaling ambiguity in such expressions. This is resolved, say,  by considering the hyperboloids $\sigma^2\equiv X^2-T^2 =2ax =$ constant and treating the light cone as the degenerate limit $\sigma\to0$ of the hyperboloids. We set $\ell_a= \nabla_a\sigma^2\propto\nabla_a x$ and take the corresponding limit.
 The motivation for this choice will become clearer later on.}
\begin{equation}
 \mathcal{H}_m[\ell_a]\equiv \frac{dQ_{m}}{\sqrt{\gamma}d^{2}xd\lambda}=T_{ab} \ell^a\ell^b
\label{hmatter} 
\end{equation}
can indeed be interpreted  as the heat density (energy per unit area per unit affine time) of the null surface, contributed by matter crossing a local Rindler horizon, as interpreted by the local Rindler observer. This interpretation works in the LRF irrespective of the nature of $T^a_b$.  So, the  need to work with $\mathcal{H}_m$, forced on us by our guiding principle, \textit{leads to} the introduction of local Rindler observers in order to interpret this quantity as the heat density.

There is an alternative interpretation of $\mathcal{H}_m$ which will prove to be useful.
Since the parameter $\lambda $ (defined through $\ell^a = dx^a/d\lambda$) is similar to a time coordinate,
we can also think of $\mathcal{H}_m[\ell_a]$ in \eq{hmatter} as the heat generated per unit area of null surface per unit time. But since there are null surfaces through any event in the spacetime, we will always have observers who see the matter heating up these surfaces! This is something we probably do \textit{not} want and we will see later on that gravity comes to our rescue. As we will see,  contribution to the heating from the microscopic degrees of freedom of the spacetime precisely cancels out $\mathcal{H}_m[\ell_a]$ on all  the null surfaces.\footnote{Our use of LRF is \textit{strictly limited} to the purpose of interpreting $\mathcal{H}_m$. I do \textit{not} introduce the notion of entropy for the Rindler horizon (as proportional to its area) or work with its variation.}

\section{Heat density of the atoms of space}

Let us get back to the task of constructing an extremum principle from which we can obtain the field equations. We have argued that the matter sector appears in the extremum principle through the combination $\mathcal{H}_m=T_{ab}(x) \ell^a \ell^b$  which has the interpretation of the heat density (or the heating rate) contributed to a null surface by the matter crossing it. We also saw that we cannot vary the metric in the extremum principle. But in any  variational principle  constructed from 
$\mathcal{H}_m$, we now have the option of varying $\ell_a$ and leaving the metric alone. Such a variational principle can take the form:
\begin{equation}
Q_{\rm tot}\equiv \int \sqrt{\gamma}\, d^2x \, d\lambda\, \left(\mathcal{H}_m[x^i, \ell_a]+\mathcal{H}_g[x^i, \ell_a]\right); \qquad \mathcal{H}_m[x^i, \ell_a] \equiv T_{ab}(x) \ell^a \ell^b
\label{Qtot1}
\end{equation} 
where we will interpret $\mathcal{H}_g$ as  the contribution to the heat density from the microscopic degrees of freedom of geometry  (`atoms of space'; I will use these two phrases interchangeably).  This should depend on both $x^i$ and $\ell_a$ for the variational principle to be well defined. 
 The success of this approach depends on our  coming up with a  candidate for $\mathcal{H}_g[x^i, \ell_a]$ which is physically well-motivated and also depends on the null vectors at each event. I will now show how this can be achieved. 

Since $\mathcal{H}_g$ has the dimension of energy density, it is convenient to write it as $L_P^{-4}f(x^i, \ell_a)$ where we have introduced a length scale $L_P$ to set the dimensions and $f$ denotes the number of atoms of space at an event $x^i$ with an extra attribute characterized by a null vector $\ell_a$. So the total heat density now becomes:
\begin{equation}
Q_{\rm tot}\equiv Q_{\rm m}+Q_{\rm g}
\equiv
\int \sqrt{\gamma}\, d^2x \, d\lambda\, \left(T^a_b(x) \ell_a\ell^b +  L_P^{-4}f[x^i, \ell_a]\right)
\label{Qtot2}
\end{equation} 
Our next task is to determine $f$.
It appears natural to assume that the number of atoms of space, $f$, (\textit{i.e.}, the microscopic degrees of freedom, contributing to the heat density) at an event $\mathcal{P}$ should be proportional to either the area or volume (which are the most primitive constructs) we can ``associate with'' the event $\mathcal{P}$. 
So we need to give precise meaning to the phrase, ``area or volume associated with'' the event $\mathcal{P}$. To do this we first introduce the notion of equi-geodesic surface,  
which can be done either in 
the Euclidean sector or in the Lorentzian sector;  we will work in the Euclidean sector. 
An equi-geodesic surface $\mathcal{S}$ is the set of all events at the same geodesic distance $\sigma$ from some event, which we take to be the origin \cite{D1,D4,D5,D6}.  A natural coordinate system to describe such a surface is given by $(\sigma, \theta_1, \theta_2, \theta_3)$ where $\sigma$, the geodesic distance from the origin, acts as the ``radial'' coordinate and $\theta_\alpha$ are the angular coordinates on the 
equi-geodesic surfaces corresponding to $\sigma =$ constant. 
The metric in this coordinate system becomes: 
\begin{equation}
 ds^2_E = d\sigma^2 + h_{\alpha\beta} dx^\alpha dx^\beta 
\label{sync}
\end{equation} 
where $h_{\alpha\beta} $ is the induced metric\footnote{This is the analogue of the synchronous frame in the Lorentzian spacetime, with $x^\alpha$ being the angular coordinates.} on $\mathcal{S}$.
The most basic quantities we can now introduce  are the volume element $\sqrt{g}\, d^4x$ in the bulk, and the area element for $\mathcal{S}$ given by $\sqrt{h}\, d^3x$. For the metric in \eq{sync}, $\sqrt{g} = \sqrt{h}$, and hence, both these  measures are identical. Using standard differential geometry, one can show \cite{D8} that, in the limit of $\sigma \to 0$, either of these  is given by: 
\begin{align}
\sqrt{h}= \sqrt{g}=\sigma^3\left(1-\frac{1}{6}\mathcal{E}\sigma ^{2}\right)\sqrt{h_\Omega}; 
\quad \mathcal{E}\equiv R^a_bn_an^b
\label{gh}
\end{align}
where $n_a=\nabla_a\sigma$ is the normal to $\mathcal{S}$ and $\sqrt{h_\Omega}$ arises from the standard metric determinant of the angular part of a unit sphere.  The second term containing  $\mathcal{E}$ gives the curvature correction to the area of (or the volume enclosed by) an equi-geodesic surface. \eq{gh} is a  standard result in differential geometry and is often presented as a measure of the curvature at any event. 

We can now  ``associate'' an area (or volume) with an event $P$ in a natural way  by the following limiting procedure:
(i) Construct an equi-geodesic surface $\mathcal{S}$ centered on $P$ at a geodesic distance $\sigma$; (ii) compute the volume enclosed by $\mathcal{S}$ and the surface area of $\mathcal{S}$; and (iii) take the limit of $\sigma\to0$ to define the area (or volume) associated with $P$.
However, as we can readily see from \eq{gh}, these measures  vanish in the limit of $\sigma \to 0$.

This is, however,  to be expected. The existence of microscopic degrees of freedom requires some form of discrete structure in the spacetime and they cannot be meaningfully defined if  the spacetime is treated as a continuum all the way, just as we cannot have a finite number of molecules in a fluid if it is treated as a continuum all the way.  Classical differential geometry, which is what we have used so far, knows nothing about any discrete spacetime structure and hence cannot give us a nonzero $\mathcal{H}_g$. To obtain a nonzero $\mathcal{H}_g$ from the above considerations, we need to ask how the geodesic interval and the spacetime metric get modified in a quantum description of spacetime and whether such a modified description will have a $\sqrt{h}$ (or $\sqrt{g}$) which does not vanish in the coincidence limit. I will now turn to this task.

\section{Points with finite area!}

There is a fair amount of evidence (see e.g., \cite{D2a,D2b,D2c,D2d,D2e,D2f}) to suggest that a primary effect of quantum gravity will be to introduce into the spacetime  a zero-point length, by modifying the geodesic interval $\sigma^2(x,x')$ between any two events $x$ and $x'$ (in a Euclidean spacetime) to a form like $\sigma^2 \to \sigma^2 + L_0^2$ where $L_0$ is a length scale of the order of Planck length.\footnote{A more general modification can take the form of $\sigma^2 \to S(\sigma^2)$  where the function $S(\sigma^2)$ satisfies the constraint $S(0) = L_0^2$ with $S'(0)$ finite. Our results  are  insensitive to the explicit functional form of  $S(\sigma^2)$. So, for the sake of  illustration, we will use $S(\sigma^2) = \sigma^2 + L_0^2$.} 

 While we do not know how quantum gravity modifies the classical metric, we  have an indirect  handle on it if we assume that quantum gravity introduces a zero point length to the spacetime. This arises as follows:
Just as the original $\sigma^2$ can be obtained from the original metric $g_{ab}$, we would expect the quantum gravity-corrected geodesic interval $S(\sigma^2)$ to arise from a corresponding quantum gravity-corrected metric \cite{D1}, which we will call the q-metric $q_{ab}$. But no such local, non-singular $q_{ab}$ can exist because, for any such $q_{ab}$, the resulting geodesic interval will vanish in the coincidence limit,  by definition of the integral. Therefore, we expect $q_{ab}(x,x')$ to be a bitensor, which should be singular at all events in the coincidence~limit $x\to x'$.
We can determine \cite{D4,D5} its form by using two~conditions: 
(i) It should lead to a geodesic interval $S(\sigma^2)$ and
(ii) a Green's function describing small metric perturbations should have a finite coincidence limit.   
(Alternatively, we can use the criterion that the flat metric  should lead to a flat q-metric.)
These conditions determine $q_{ab}$ uniquely \cite{D5} in terms of $g_{ab}$ (and its associated geodesic interval $\sigma^2$). 
We find that: 
\begin{align}
q_{ab}=Ah_{ab}+ B n_{a}n_{b};\qquad q^{ab}=\frac{1}{A}h^{ab}+\frac{1}{B}n^{a}n^{b}
\label{qab}
\end{align}
with
\begin{align}
B=\frac{\sigma ^{2}}{\sigma ^{2}+L_{0}^{2}};\qquad A=\left(\frac{\Delta}{\Delta _{S}}\right)^{2/D_{1}}\frac{\sigma ^{2}+L_{0}^{2}}{\sigma ^{2}};\qquad n_a=\nabla_a\sigma
\label{defns}
\end{align}
where $D$ is the  spacetime dimension , $D_k\equiv D-k$.
The $\Delta$ is the Van-Vleck determinant related to the geodesic interval $\sigma^2 $ by:
\begin{align}
\Delta (x,x')=\frac{1}{\sqrt{g(x)g(x')}}\textrm{det}\left\lbrace \frac{1}{2}\nabla _{a}^{x}\nabla _{b}^{x'}\sigma ^{2}(x,x') \right\rbrace
\end{align}
and $\Delta_S$ is the corresponding quantity computed by replacing  $\sigma^{2}$  by $S(\sigma^{2})$
(and $g_{ab}$ by $q_{ab}$ in the covariant derivatives)
 in the above definition.
For our purpose of determining $\mathcal{H}_g$, we will 
compute the area element ($\sqrt{h}\, d^3 x$) of an equi-geodesic surface and the volume element ($\sqrt{q} \ d^4x$) for the region enclosed by it, 
using the renormalized q-metric.  (For the q-metric in \eq{qab}, corresponding to the $g_{ab}$ in \eq{sync}, these two measures will not be equal, because $q_{00} \neq 1$.)
If our ideas are correct,  we should get a non-zero limit and there must be a valid mathematical reason to prefer one of these measures over the other. 

The computation is 
straightforward  and we find  (for \mbox{$S(\sigma^2)=\sigma^2+L_0^2$} in $D=4$, though similar results \cite{paperD,D7} hold in the more general case in $D$ dimensions) that:
\begin{align}
\sqrt{q}=\sigma \left(\sigma ^{2}+L_{0}^{2}\right)\left[1-\frac{1}{6}\mathcal{E}\left(\sigma ^{2}+L_{0}^{2}\right)\right]\sqrt{h_\Omega}
\label{qfinal}
\end{align}
and:\footnote{This result is
rather subtle. One might think that the result in \eq{hfinal} (which is  $\sqrt{h}=A^{3/2}\sqrt{g}$)  arises from the standard result  \eq{gh}, by the simple replacement of $\sigma^2\to(\sigma^2+L_{0}^{2})$. But note that this replacement does \textit{not} work for the result in \eq{qfinal} (which is $\sqrt{q}=\sqrt{B}A^{3/2}\sqrt{g}$) due to the $\sqrt{B}=\sigma(\sigma ^{2}+L_{0}^{2})^{-1/2}$ factor which has the limiting form $\sqrt{B}\approx\sigma/L_{0}$ when $\sigma\to0$. This is why  each event has zero volume, but finite area associated with it!. A possible insight into this curious feature is provided by the following fact:
The leading order dependence of $\sqrt{q}d\sigma\approx\sigma d\sigma$ shows that the volumes scale as $\sigma^2$ while the area measure is finite. This, in turn, is related to the fact \cite{paperD} that \textit{the effective dimension of the renormalized spacetime becomes $D=2$ close to Planck scales,} independent of the original $D$. This  result  has been noticed by several people (\cite{z1,z2,z3,z4}; also see \cite{nicolini})
in different, but specific, models of quantum gravity. Our approach leads to this result in a \textit{model-independent} manner, which, in turn, is connected with the result that events have zero volume, but non-zero area.}
\begin{align}
\sqrt{h}=\left(\sigma ^{2}+L_{0}^{2}\right)^{3/2}\left[1-\frac{1}{6}\mathcal{E}\left(\sigma ^{2}+L_{0}^{2}\right)\right]\sqrt{h_\Omega}
\label{hfinal}
\end{align}
When $L_{0}^{2}\to0$, we recover the standard result in \eq{gh}, as we should. Our interest, however, is in the limit $\sigma^2\to0$ at finite $L_0$.
Something remarkable happens when we take this limit. The volume measure $\sqrt{q}$ vanishes (just as for the original metric) but $\sqrt{h}$ has a non-zero limit:
\begin{align}
\sqrt{h}= L_{0}^{3}\left[1-\frac{1}{6}\mathcal{E}L_{0}^{2}\right]\sqrt{h_\Omega}
\label{hlimit}
\end{align}
The q-metric (which we interpret as representing the renormalized/dressed spacetime)
attributes to every point in the spacetime a finite area measure, but a zero volume measure! 
Since $L_0^3\sqrt{h_\Omega}$ is the volume measure of the $\sigma=L_0$ surface, we define \cite{tpentropy} the dimensionless density of the atoms of spacetime, contributing to the gravitational heat density, as:
\begin{equation}
f(x^i,n_a)\equiv \frac{\sqrt{h}}{L_0^3\sqrt{h_\Omega}} =1-\frac{1}{6}\mathcal{E}L_{0}^{2}
=1-\frac{1}{6} L_{0}^{2} R_{ab}n^an^b
\label{denast}
\end{equation}

\begin{figure}
 \begin{center}
  \includegraphics[scale=0.63]{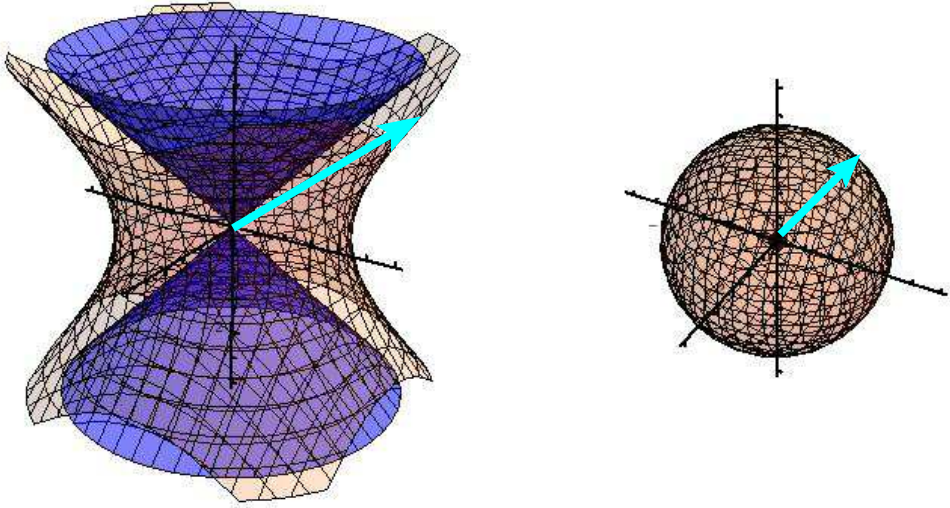}
 \end{center}
\caption{(a) Left: In the local inertial frame, the light cones originating from an event (taken to be the origin) are the null surfaces with $R^2-T^2 = 0$,  with a normal $\ell_a$. The local Rindler observers live on the hyperboloid $R^2-T^2 = \sigma^2 = $ constant around these light cones and perceive a patch of the light cone as a local Rindler horizon with a non-zero temperature. The arrow denotes the normal to the hyperbola. (b) Right: In the Euclidean sector, the hyperboloid $R^2-T^2 = \sigma^2$ will become a sphere $R^2 + T_E^2 = \sigma_E^2$ and the  normal to the hyperboloid  becomes the normal to the sphere. The light cone $R^2 - T^2 =0$ will go over to $R^2 + T_E^2 =0$ which collapses into the origin.  The limit $\sigma_E \to 0$ in the Euclidean sector corresponds to approaching the Rindler horizon in the Lorentzian sector. In this limit  the hyperboloid degenerates into the light cones emanating from $\mathcal{P}$. The direction of the normal to the sphere becomes ill-defined in the Euclidean sector when the radius of the sphere shrinks to zero. In the Lorentzian sector, we can take it to be the normal to the null surface in the limit when the hyperboloid degenerates to the light cone.
 The dependence of $f$ on $n_a$ in the Euclidean equi-geodesic surface is what translates into its dependence on the null normal $\ell_a$ in the Lorentzian sector.}
\label{fig:lightcones}
\end{figure}

So far we have been working in the Euclidean sector with $n_a=\nabla_a\sigma$ being normal to the equi-geodesic surface. The limit $\sigma\to0$ in the Euclidean sector makes the equi-geodesic surface shrink to the origin. \textit{But, in the Lorentzian sector this leads to the null surface which acts as the local Rindler horizon around this event.} So, in this limit, we can identify $n_a$ with the normal to the null surface $\ell_a$ and express the gravitational heat density as
\begin{equation}
f(x^i,\ell_a) 
=1-\frac{1}{6} L_{0}^{2} R_{ab}\ell^a\ell^b
\label{denastx}
\end{equation}
To see this in some detail, consider the Euclidean version of the local Rindler frame.\footnote{There are two ways of extending the null surface and the Rindler observers off the $TX$ plane. One can extend the null line (45 degree line in Fig.~\ref{fig:daviesunruh}) to the null \textit{plane}  $T=X$ in spacetime and similarly for the hyperboloid. Alternatively, one can 
extend the null line  to the null \textit{cone} by $R^2-T^2=0$ with $R^2=X^2+Y^2+Z^2$ and the hyperboloid $R^2-T^2=$ constant will go `around' the null cone in the spacetime (see the left part of Fig.~\ref{fig:lightcones}). Observers living on this hyperboloid will use their respective (rotated) $X$ axis to study the Rindler frame physics.} The local Rindler observers, living on the hyperboloid $R^2 - T^2 = \sigma^2$ perceive local patches of the light cone $R^2 - T^2 =0$ as their horizon (see the left half of Fig.~\ref{fig:lightcones}).
If we now analytically continue to the Euclidean sector, the hyperboloid $R^2 - T^2 = \sigma^2$ will become a sphere $R^2 + T_E^2 = \sigma_E^2$ (see the right half of Fig.~\ref{fig:lightcones}). The light cones $R^2 - T^2 =0$ goes over to $R^2 + T_E^2 =0$ and collapses into the origin. So, taking the limit $\sigma_E \to 0$ in the Euclidean sector corresponds to approaching the local Rindler horizons in the Lorentzian sector. This is the limit in which the hyperboloid degenerates into the light cones emanating from $\mathcal{P}$. The normal $n_a$ to the Euclidean sphere can then be identified with the normal to the null surface $\ell_a$. The dependence of $f$ on $n_a$ in the Euclidean equi-geodesic surface is what translates into its dependence on the null normal $\ell_a$ in the Lorentzian sector.

So the contribution to the \textit{gravitational} heat density on a null surface in \eq{Qtot2} is obtained by integrating $L_P^{-4}f(x^i,\ell_j)$ over the volume:
\begin{equation}
Q_{g}=\int\frac{\sqrt{\gamma}d^2xd\lambda}{L_P^4}f(x^i,\ell_j) =
\int\frac{\sqrt{\gamma}d^2xd\lambda}{L_P^4}\left[1-\frac{1}{6}L_{0}^{2}(R_{ab}\ell^a\ell^b)\right]
\label{corres}
\end{equation} 
The numerical factor in front of the second term depends on the  ratio $L_0/L_P$ which we expect to be of order unity, if $L_P$ is the standard Planck length. This ratio  cannot be determined without better knowledge of quantum gravity. But we get the correct equations with the choice $L_0^2=(3/4\pi)L_P^2$, which we will make.\footnote{If we assume that the gravitational heat density is $\mu f/L_P^4$ where $\mu$ is a factor of order unity, the numerical coefficients will change. We take $\mu=1$ as a natural choice.}
The full variational principle is then based on the functional:
\begin{eqnarray}
Q_{\rm tot}&\equiv& \int \sqrt{\gamma}\, d^2x \, d\lambda\, (T^a_b(x) \ell_a\ell^b+L_P^{-4}f[x^i, \ell_a])\nonumber\\
&=&\int \sqrt{\gamma}\, d^2x \, d\lambda\,\left[  
\frac{1}{L_P^4}+\left\{T^a_b-\frac{1}{8\pi L_P^2}R^a_b\right\}\ell_a\ell^b
\right]
\label{Qtot3}
\end{eqnarray} 
Extremising this functional with respect to $\ell_a$ after introducing a Lagrange multiplier to keep $\ell^2=0$, and demanding the extremum holds for all $\ell_a$ at an event again leads to the result $R^a_b - (8\pi L_P^2) T^a_b = \lambda(x) \delta^a_b$ where $\lambda(x)$ is the Lagrange multiplier. Using the  Bianchi identity and $\nabla_a T^a_b =0$, we will recover the field equations except for a cosmological constant. (There is actually a way to determine its value, which I will discuss in Sec. \ref{sec:cc}.)

There are several points which are noteworthy about this result which I will now comment upon:

\subsection{A crucial minus sign}
We defined the heat density of gravity as $L_P^{-4}f(x,n_a)$ with the number of atoms of space $f$ being given by the limit:
\begin{equation}
f(x^i,n_a)\equiv \lim_{\sigma\to 0}  \frac{\sqrt{h(x,\sigma)}}{L_0^3\sqrt{h_\Omega}}
\label{denast1}
\end{equation}
in a renormalized spacetime with a zero-point length. The result had the combination $R_a^bn_bn^a$, at the relevant order, which is crucial. Further, this term came with a \textit{minus sign}  without which the programme would have failed. 

These results bring to the center-stage the geodesic interval $\sigma^2(x,x')$ (rather than the metric) as the proper variable to describe spacetime geometry \cite{D7}. In a classical spacetime,  $\sigma^2(x,x')$ and $g_{ab}(x)$ contain the same amount of information and each is derivable from the other. But $\sigma^2(x,x')$ seems to be better suited to take into account quantum gravitational effects to a certain extent.

\subsection{Dissipation without dissipation}
\label{sec:dwd}

 We mentioned earlier that one could have also thought of  $\mathcal{H}_m$ as the heating \textit{rate} of the null surface by matter. We see that the corresponding  heating rate of the null surface by  the atoms of space precisely compensates this on-shell leaving an unobservable constant factor $L_P^{-4}$. (Since we can add any   quantity  independent of $\ell_a$ to the integrand of $Q_{tot}$, we can even renormalize this away; we leave it there because it has an interesting implication for the \cc ; see Sec. \ref{sec:cc}). So the field equation, which is now in the form of \eq{Gnn}, has a clear physical meaning; both sides represent the  heating rate of null surfaces and Einstein's equation arises as  a heat balance equation.

This interpretation is reinforced by the fact that the term $R_{ab}\ell^a\ell^b$ is related to the ``dissipation without dissipation'' \cite{sanvedtp} of the null surfaces, which can be described as follows: Introduce the second null vector $k^a$ and define the 2-metric on the cross-section of the null surface in the standard manner, $q_{ab}=g_{ab}+k_ak_b+\ell_a\ell_b$. Next define the expansion $\theta\equiv\nabla_a\ell^a$ and shear $\sigma_{ab}\equiv \theta_{ab}-(1/2)q_{ab}\theta$ where $\theta_{ab}=q^i_aq^j_b\nabla_i\ell_j$. (We take the null congruence  to be affinely parametrized.) One can then show that (see e.g., eq(A60) of \cite{A19}):
\begin{equation}
-\frac{1}{8\pi L_P^2}R_{ab}\ell^a\ell^b=\left[2\eta \sigma_{ab}\sigma^{ab}+\zeta\theta^2\right]
+ \frac{1}{8\pi L_P^2}\frac{1}{\sqrt{\gamma}}\frac{d}{d\lambda}(\sqrt{\gamma}\theta)\equiv\mathcal{D}+\frac{1}{8\pi L_P^2}\frac{1}{\sqrt{\gamma}}\frac{d}{d\lambda}(\sqrt{\gamma}\theta)
\label{rai}
\end{equation} 
where $\eta=1/16\pi L_P^2,\zeta=-1/16\pi L_P^2$ are the shear and bulk viscous coefficients \cite{A26,A27,membrane} and $\mathcal{D}$ is the viscous dissipation rate. So, ignoring the total divergence term and the constant, the relevant part of $Q_{tot}$ in \eq{Qtot3} can be expressed as
\begin{equation}
Q_{\rm tot}
=\int \sqrt{\gamma}\, d^2x \, d\lambda\, \left(T^a_b(x) \ell_a\ell^b +\left[2\eta \sigma_{ab}\sigma^{ab}+\zeta\theta^2\right]\right)
=\int \sqrt{\gamma}\, d^2x \, d\lambda\, \left(T^a_b(x) \ell_a\ell^b +\mathcal{D}\right)
\label{Qtoty}
\end{equation}
Both terms now have an interpretation of the rate of heating (due to matter or atoms of spacetime).\footnote{Equation (\ref{rai}) is just a restatement of the Raychaudhuri equation. What is relevant in the extremum principle are the \textit{quadratic} terms in shear and expansion, while the term giving the  change in the cross-sectional area of the congruence is a total divergence and is irrelevant. This tells us that ignoring the quadratic terms of the Raychaudhuri equation can miss a key element of physics;  see also \cite{dawood1}.} Our extremum principle can be thought of extremising the rate of production of heat on the null surface.

\subsection{Connection with the entanglement entropy}

The combination $R^a_b n_an^b$, which is crucial to our results, arises in several other geometrical quantities just as it came up in the area measure $\sqrt{h}$ in \eq{hlimit}. In particular, this combination has some relevance to entanglement entropy, in the following sense. In the literature, it is usually claimed that the entanglement entropy  $S$ of a field partitioned by, say, a horizon is proportional to the area of the horizon. \textit{To be precise, this statement is  meaningless because  the entanglement entropy $S$ is a divergent quantity.} For example, for  a free, massless, scalar field in $D$ dimensional  spacetime, $S$ is given by the expression:
\begin{equation}
 S=\frac{A_{D-2}}{12}\int_0^\infty \frac{ds}{s}K_{D-2}(x,x;s)
\label{SfromK}
\end{equation} 
where $A_{D-2}$ is the transverse area   and $K(x,y;s)$ is the Schwinger proper time Kernel \cite{kernela,kernelb}. In the coincidence limit, 
the  $K_{D-2}(x,x;s)\propto s^{-(D-2)/2}$ and the integral in \eq{SfromK} diverges as $L_{c}^{-(D-2)}$ at the lower limit, where $L_{c}$ is a lower cutoff length scale. In $D=4$ spacetime, this gives $S\propto A_\perp/L_{c}^2$, which diverges quadratically, which is the standard result. 

The introduction of a zero-point length into the spacetime corresponds to introducing a regulator $\exp(-L/s)$ into the heat Kernel (with some cutoff scale $L$) \textit{which will render the entanglement entropy finite} \cite{tpentangle} and these calculations meaningful. What is more, the Kernel will also pick up a Van Vleck determinant in a curved spacetime which also contains the factor $R^a_b n_an^b$ over and above the flat spacetime result. \textit{This suggests that one may be able to relate $f(x^i, n_a)$ in \eq{denast} to the entanglement entropy which could provide an alternative interpretation to our results.}
This idea could be tested by  computing  the entanglement entropy of a field in the renormalized spacetime with zero-point length, which renders it finite in a systematic manner.

\subsection{The mechanism that couples matter to spacetime}

The coupling between spacetime geometry and matter is now through a vector field $\ell_a$ and --- at the lowest order ---  they  couple to $\ell_a$ through the terms $R^a_b \ell_a \ell^b$ and $T^a_b \ell_a \ell^b$ respectively. The physical origin of these  two couplings is quite distinct.  The $T^a_b \ell_a \ell^b$ arises from the behaviour of matter crossing the local Rindler horizon and the $\ell_a$ in this expression represents the normal to the local Rindler horizon. The $R^a_b \ell_a \ell^b$ term, however, arose from the limit of the area measure $\sqrt{h}$ in a spacetime with a zero-point length. The $n_a$ in this case is identified with the normal $\ell_a$ to the null surface through a limiting process when one takes the limit $\sigma \to 0$ in the Euclidean sector. This, in turn, depends  on the fact that the condition $\sigma^2 (x,y) =0$ will lead to $x=y$ in the Euclidean space while it will describe all events connected by a null ray in the Lorentzian space.

More generally, we can introduce a vector $n_a $ that couples to spacetime curvature through $R^a_b n_a n^b$ and to matter through $T^a_b n_a n^b$, thereby providing an indirect coupling between matter and geometry. This procedure is conceptually rather satisfactory. In the conventional approach to Einstein's theory we actually do not have a \textit{mechanism} which tells us how $T^a_b$ ends up curving the spacetime. The relation  $G^a_b = 8\pi T^a_b$  equates apples and oranges; the left hand side is purely geometrical while the right hand side is made of  matter with a large number of discrete (quantum) degrees of freedom. An equation of the kind, $G^a_bn_a n^b = 8\pi T^a_bn_a n^b$, on the other hand, does better in this regard.  We can hope to interpret both sides independently (say,  as the heat densities) and think of this equation as a balancing act performed by spacetime. Such a description is reinforced by the extremum principle in \eq{Qtot3} in which both $R^a_b n_a n^b$ and $ T^a_bn_a n^b$ can be thought of as distorting the value of $f$ from unity, and the gravitational field equations restore the value $f=1$ on-shell. (I will say a little more about this in the last section.)

Since  classical gravity is obtained by extremising $Q_{\rm tot}$ in \eq{Qtot3} with respect to $n_a$, we could ask whether it is meaningful to attempt a quantum theory for $n_a$ using a path integral. This would require evaluating the path integral 
\begin{equation}
 Z= \int \mathcal{D} n_a \, \delta \left[ n^2\right] \, \exp i Q_{\rm tot} \left[ n_a\right]  
\end{equation}
Incorporating  $\delta \left[ n^2\right]$ through a functional Fourier transform with respect to a Lagrange multiplier  field $\lambda [x]$, we can reduce this to the form:
\begin{equation}
Z \propto \prod_x \int \mathcal{D} \lambda(x)  \int \mathcal{D} n_a(x) \exp i \left[ (M^a_b + \lambda \delta^a_b)n_a n^b \right]_x  
\propto  \prod_x\int \mathcal{D} \lambda(x)\; {\rm Det}^{-1/2} \left[ M^a_b + \lambda  \delta^a_b \right]_x 
\end{equation}
where  $M^a_b \equiv T^a_b - (8\pi L_P^2)^{-1} R^a_b$. This is a rather intriguing theory which will lead to Einstein's equations in the saddle point limit if we demand that the saddle point condition should hold for all $n_a$.  The Euclidean version of this theory suggests working with $\exp(Q_g)$ and it will be interesting to explore this further.\footnote{If we write the null vector as $\ell_a\equiv q(x)\bar\ell_a$ where $\bar\ell_a\equiv\nabla_a\sigma$ is an affinely parametrized null normal, then we can upgrade just $q(x)$ to a dynamical field with the Lagrangian $L=(1/2)(\partial q)^2-q^2[M^a_b\ell_a\ell^b]$.  Integrating over $q(x)$, treating it as a rapidly varying fluctuation (with $[M^a_b\ell_a\ell^b]$ approximately constant) will lead  to an  effective action $A_{\rm eff}$ for $[M^a_b\ell_a\ell^b]$ such that $\delta A_{\rm eff}=0$ is the same as $\delta Q_{\rm tot}=0$. I will describe this in detail elsewhere.}

\subsection{Solution to the \cc\ problem}
\label{sec:cc}

On shell, when the equations of motion hold, the two terms in the curly brackets in  \eq{Qtot3} cancel each other and the net heat density has the Planckian value $1/L_P^4$, which, of course, has no gravitational effect. But it tells us that there \textit{is} a zero-point contribution to the degrees of freedom in spacetime, which, in dimensionless form, is just unity. Therefore, it makes sense to ascribe $A/L_P^2$ degrees of freedom to an area $A$, which is consistent with what we know from  earlier results in this subject.
So a two-sphere $S^{(2)}$of radius $L_P$ has $f_{S^{(2)}}=4\pi L_P^2/L_P^2=4\pi$, which  was the crucial input that was used in a previous work to determine the numerical value of the \cc\ for our universe. (This is similar to assigning $dN/d^3xd^3p=f(x^i,p_a)$ molecules to a phase volume. In kinetic theory, we do not worry about the fact that $f$ is not always an integer. In the same spirit, we are not concerned by the fact that $4\pi$ is not an integer.). Thus, the microscopic description does allow us to determine \cite{C8,C9} the value of the \cc, (which arose as an integration~constant), as it should in any complete description.

Let me elaborate a little bit on this aspect, since it can provide a solution to what is usually considered the most challenging problem of theoretical physics today. 

Observations suggest that our universe  has three distinct phases of expansion: (i) An inflationary phase with an approximately constant density $\rho_{inf}$, fairly early on. (ii) A phase dominated, first by radiation and then by matter, with $\rho=\rho_{eq}[x^{-4}+x^{-3}]$, where  $a_{eq}$ is the epoch at which the matter and radiation densities were equal, $x(t)\equiv a(t)/a_{\rm eq}$ and the $\rho_{eq}$ is a  constant equal to the energy density of either matter or radiation at $a = a_{\rm eq}$. (iii) An accelerated phase of expansion at late stages, driven by the energy density $\rho_\Lambda$ of the cosmological constant. These three constants $[\rho_{inf},\rho_{eq},\rho_\Lambda]$  completely specify the dynamics of our universe and act as its signature. Of these, $\rho_{inf}$ and $\rho_{eq}$ can, in principle, be determined by standard high energy physics. But we need a \textit{new} principle to fix the value of $\rho_\Lambda$, which is related to the integration constant that appears in the field equations in our approach.

It turns out that such a universe, with these three phases, harbors a new \textit{conserved} quantity, which is the number $N$ of length scales (or radial geodesics), that cross the Hubble radius during any of these phases \cite{C8,C9}. Any physical principle which can determine the value of $N$ during the radiation-matter dominated phase, say, will fix the value of $\rho_\Lambda$ in terms of $[\rho_{inf},\rho_{eq}]$. Taking the modes into the very early phase, we can fix the value of this conserved quantity $N$  at the Planck scale, as the degrees of freedom in a two-sphere of radius $L_P$. In other words, we take $N=4\pi L_P^2/L_P^2 = 4\pi$. This, in turn, leads to a remarkable prediction relating the three densities~\cite{C8,C9}:
\begin{equation}
 \rho_\Lambda \approx \frac{4}{27} \frac{\rho_{\rm inf}^{3/2}}{\rho_{\rm eq}^{1/2}} \exp (- 36\pi^2)
\label{ll6}
 \end{equation} 
From cosmological observations, we know that $\rho_{eq}^{1/4} = (0.86 \pm 0.09) \text{eV}$; if we assume that the range of the inflationary energy scale is $\rho_{\rm inf}^{1/4} = (1.084-1.241) \times 10^{15}$ GeV, we get $\rho_{\Lambda} L_P^4 = (1.204 - 1.500) \times 10^{-123}$, which is consistent with the observations! 
This  approach for solving the cosmological constant problem provides a unified view of cosmic evolution, connecting the three phases through \eq{ll6} in contrast with standard cosmology in which the three phases are joined together in an unrelated, \textit{ad hoc} manner.

Moreover, this approach to the cosmological constant problem \textit{makes a falsifiable prediction}, unlike any other approach I am aware of. From the observed values of $\rho_\Lambda$ and $\rho_{\rm eq}$, we can \textit{predict} the energy scale of inflation within a very narrow band --- to within a factor of about five --- if we include the ambiguities associated with reheating. If future observations show that inflation occurred at energy scales outside the band of $(1-5)\times 10^{15}$ GeV, our model for explaining the value of \cc\ is ruled out.

\section{Summary of the logical structure}

Let me reiterate the logical sequence described in this work which leads to a completely different perspective on gravity and the derivation of its field equations.

\begin{itemize}

 \item[$\blacktriangleright$] We postulate that the gravitational field equations should arise from an extremum principle which remains invariant under the transformation $T^a_b \to T^a_b + \delta^a_b $(constant).

\item[$\blacktriangleright$] This leads to two conclusions: (a) The metric tensor cannot be a dynamical variable which is varied in the extremum principle.
(b) The $T^a_b$ should appear in the extremum principle though  the combination $\mathcal{H}_m(x^i, \ell_a)\equiv T^a_b \ell_a \ell^b$ where $\ell_a$ is a null vector.

\item[$\blacktriangleright$] We next look for a physical interpretation of $\mathcal{H}_m(x^i, \ell_a)$ for an arbitrary  $T^a_b$ and find that, in any spacetime,  the local Rindler observers will interpret $\mathcal{H}_m(x^i, \ell_a)$ as the heat density contributed to a null surface by the matter crossing it. This interpretation works for \textit{any} $T^a_b$ and provides a strong \textit{motivation} for introducing local Rindler observers in the spacetime. 

\item[$\blacktriangleright$] Since (i) the metric cannot be a dynamical variable and (ii) we now have the auxiliary null vector field $\ell_a$ arising through $\mathcal{H}_m(x^i, \ell_a)$, we look for an extremum principle in which $\ell_a$ is varied. The extremum should hold for all null vectors at any event and constrain the background metric. 

\item[$\blacktriangleright$] We take the integrand of the extremum principle  to be  $\mathcal{H}_m(x^i, \ell_a) + \mathcal{H}_g(x^i, \ell_a)$  where $\mathcal{H}_g(x^i, \ell_a) \equiv L_P^{-4}f(x^i, \ell_a)$ is interpreted as the heat density due to the microscopic spacetime degrees of freedom. We have introduced a length scale $L_P$ from dimensional considerations and $f$ is the dimensionless count of the microscopic degrees of freedom.

\item[$\blacktriangleright$]  A discrete count for  the microscopic degrees of freedom implies a discrete nature for spacetime at Planck scales. We incorporate this fact by an effective, renormalized/dressed metric $q_{ab}$ which ensures that the geodesic distance of the effective metric has a zero-point length and is given by $\sigma^2 (x, x') + L_0^2$ where $\sigma^2 (x, x')$ is the geodesic distance of the classical spacetime. We take $L_0^2=(3/4\pi) L_P^2$.

\item[$\blacktriangleright$] The number of spacetime degrees of freedom, at an event, is taken to be proportional to the area of an equi-geodesic surface centered at that event in the limit of vanishing geodesic distance  (see \eq{denast1}). With this choice, one obtains an extremum principle based on \eq{Qtot3}. Varying this with respect to all null vectors $\ell_a$ and demanding the equation to hold for all $\ell_a$ at an event leads to Einstein's equation with an undetermined cosmological constant.

\item[$\blacktriangleright$] When equations of motion hold, we can assign $A/L_P^2$ degrees of freedom with every area element $A$ in spacetime.  This, in turn, allows us to fix the value of the undetermined \cc\ correctly and \textit{provides a solution to the cosmological constant problem}. 

\end{itemize}

\section{Conclusions and open questions}

The approach outlined here is based on the idea that gravity is the thermodynamic limit of the statistical mechanics of certain microscopic degrees of freedom (`atoms of space'). In the  thermodynamic limit, deriving the field equations of classical gravity is algebraically straightforward --- one might even say trivial, but let us not shun simplicity! It is obtained from an extremum principle based on the functional: 
\begin{equation}
 Q_{\rm thermo}
=\int \sqrt{\gamma}\, d^2x \, d\lambda\,  
\left\{T^a_b-\frac{1}{8\pi L_P^2}R^a_b\right\}\ell_a\ell^b
\label{Qtot4}
\end{equation} 
Varying $\ell_a$ in $Q_{\rm thermo}$ with the constraint that $\ell^2=0$, and demanding that the result holds for all $\ell_a$ will lead to Einstein's equation with an arbitrary  cosmological constant. 
The approach works for a wild class of gravitational theories including the \LL\ models.

As far as classical gravity goes, that is the end of the story. But we could enquire about the physical meaning of this extremum principle.\footnote{We do \textit{not} enquire about the physical meaning of the Einstein-Hilbert action --- it has none --- in the conventional approach; but then, we are now trying to \textit{improve} on the conventional approach!} The combination $T^a_b\ell_a\ell^b$ is invariant under the shift $T^a_b \to T^a_b +\delta^a_b$ (constant) --- which was the original reason to put it in the variational principle. 
Postulating that the extremum principle must be invariant under the transformation $T^a_b \to T^a_b + \delta^a_b $(constant) naturally \textit{leads to} the introduction of $T^a_b \ell_a\ell^b$ in the variational principle and to the local Rindler observers for its interpretation.
As I mentioned earlier, this by itself is a valuable insight and shows the  connection between two features --- viz. the immunity of gravity to shifts in the cosmological constant, and the thermodynamic interpretation of gravity --- which were considered as quite distinct. 

Further, $T^a_b \ell_a\ell^b$ \textit{does} have the interpretation as the heat density of matter on any null surface. It is also  possible to interpret $R^a_b \ell_a\ell^b$ as the heating rate (``dissipation without dissipation''; see Sec. \ref{sec:dwd}) of the null surface \cite{h214}, thereby providing a purely thermodynamic underpinning for classical gravity.  None of this poses any conceptual or technical problem.

The real issues arise when we try to go beyond the classical theory and provide a semi-classical or quantum gravitational interpretation of this result. 
The description in terms of the expression in \eq{denast} is  approximate and only valid when $L_P^2 R^a_b n_an^b \lesssim 1$. The entire description, based on the q-metric acting as a proxy for the effective spacetime metric, cannot be trusted too close to Planck scales.  We expect it to capture the quantum gravitational effects to a certain extent, but at present we have no way of quantifying the accuracy of this approach. (It should also be noted that the identification $n_a\approx \ell_a+.....$ might have further corrections close to Planck scales.)

We also do not know how to deal with matter fields in a quantum spacetime and it is not clear how to introduce $T^a_b$ in a systematic way. In fact, the \textit{only} reason to vary the metric tensor in an extremum principle --- a procedure which I have argued against --- is to obtain $T^a_b$ classically and $\langle T^a_b \rangle$ semi-classically. None of the thermodynamic derivations,  which leads to the exact (rather than linearized) field equations, obtains  $T^a_b$ (or $\langle T^a_b \rangle$) from fundamental considerations based on a matter action.  It is ironic that the problem arises from the matter sector rather than from the gravity sector!

The ideas outlined in this work suggest  a more radical solution. Matter, as we understand it, is quantum mechanical; which essentially means that it is made of \textit{discrete}  degrees of freedom. How does such a \textit{discrete} structure end up curving the \textit{continuum} geometry? That is, what is the actual \textit{ mechanism} by which  $T^a_b$ produces  $G^a_b$? The approach developed here suggests that one needs to introduce certain ``hidden variables'', viz., the  auxiliary vector field $n_a$, which encodes the discrete nature of geometry, and couple it to $T^a_b$ (which is also fundamentally discrete in nature). The continuum geometry has to be related to $n_a$, at scales much larger than the Planck scale, by a suitable approximation just as the continuum density or pressure of a fluid arises when we average over the discreteness of the molecules. 
I have outlined one possible way in which this idea could be implemented, but it is by no means unique. Further exploration of this approach could lead to a better understanding of how matter \textit{really} ends up curving the spacetime. 

\section*{Acknowledgements}

My research work is partially supported by J.C.Bose Fellowship of Department of Science and Technology, India. I thank  Sumanta Chakraborty, Sunu Engineer, Dawood Kothawala and  Kinjalk Lochan for discussions and  comments on the first draft.


\begin{thebibliography}{99}
 


\bibitem{A26}  
  Damour T  1979
  {\em Th`ese de doctorat
  d’´Etat, Universit´e Paris}.

\bibitem{A27}
 Damour T   1982 ``{Surface effects in black hole physics},'' {\em Proceedings of the
  Second Marcel Grossmann Meeting on General Relativity}.

\bibitem{membrane} Thorne K. S.,  Price R. H. and MacDonald D. A. 1986 \textit{Black Holes: The Membrane Paradigm} (Yale University Press)


\bibitem{A35}
Jacobson T. 1995 ``Thermodynamics of space-time: The Einstein equation of state'',   {\em
Phys.~Rev.~Lett.}, {\bf 75}, 1260. 

\bibitem{tp02}   Padmanabhan T., 2002 ``Classical and Quantum Thermodynamics of horizons in spherically symmetric spacetimes'', \textit{Class.Quan.Grav.}, \textbf{19}, 5387. [gr-qc/0204019] 

\bibitem{tp04}  Padmanabhan T., 2004 ``Gravitational Entropy of Static Spacetimes and Microscopic Density of States'', \textit{Class.Quan.Grav.}, \textbf{21}, 4485. [gr-qc/0308070] 

\bibitem{aseemtp}  Padmanabhan T., Aseem Paranjape, 2007 ``Entropy of Null Surfaces and Dynamics of Spacetime'',\textit{ Phys.Rev. D} \textbf{75} 064004. [gr-qc/0701003] 


\bibitem{tg17} Padmanabhan T., 2010 ``Thermodynamical Aspects of Gravity: New insights,''
 {\em Rept. Prog.
  Phys.}, {\bfseries 73}  046901. arXiv:0911.5004 [gr-qc]


\bibitem{tpmpla}  Padmanabhan T., 2010 ``Equipartition of energy in the horizon degrees of freedom and the emergence of gravity'', \textit{Mod.Phys.Letts.}, \textbf{A 25}, 1129-1136. [arXiv:0912.3165] 

\bibitem{tg16} Verlinde E. P., 2011 ``{On the Origin of Gravity and the Laws of Newton},''
  {\em JHEP} {\bfseries 1104}, 029 


\bibitem{lee2}  Smolin L., ``General relativity as the equation of state of spin foam'' [arXiv:1205.5529]

\bibitem{A19}
Padmanabhan T., 2014 ``General Relativity from a Thermodynamic Perspective'',
 {\em Gen.~Rel.~Grav.}  {\bf 46}, 1673.  [arXiv:1312.3253]


\bibitem{A11}
Padmanabhan, T., 2015 ``Emergent Gravity Paradigm: Recent Progress'', {\em Mod. Phys. Lett. A},
 {\bf 30}, 1540007.


\bibitem{tg14} 
Jacobson T., ``{Entanglement equilibrium and the Einstein equation},''
[arXiv:1505.04753]


\bibitem{lee1} Smolin L., ``The thermodynamics of quantum spacetime histories'',  [arXiv:1510.03858]


\bibitem{A21}
Parattu K., B.~R. Majhi, and T.~Padmanabhan, 2013  ``The Structure of the Gravitational Action and its relation with Horizon Thermodynamics and Emergent Gravity Paradigm'' {\em Phys.
  Rev. D}, {\bfseries 87},  124011
  [arXiv:1303.1535]. 


\bibitem{A20}  Chakraborty S., T. Padmanabhan, ``Evolution of Spacetime arises due to the departure from Holographic Equipartition in all Lanczos-Lovelock Theories of Gravity'', 2014 \textit{Phys. Rev.,} \textbf{D 90}, 124017  [arXiv:1408.4679] 




\bibitem{hm258}  Chakraborty S., T. Padmanabhan, ``Thermodynamical interpretation of the geometrical variables associated with null surfaces'', 2015 \textit{Phys. Rev.}, \textbf{D 92}, 104011 [arXiv:1508.04060] 



\bibitem{h256}  Chakraborty S., Krishnamohan Parattu, T. Padmanabhan, ``Gravitational field equations near an arbitrary null surface expressed as a thermodynamic identity'', 2015 \textit{JHEP}, \textbf{10}, 097  [arXiv:1505.05297].



\bibitem{tpsurf}  Padmanabhan T., 2010 ``Surface Density of Spacetime Degrees of Freedom from Equipartition Law in theories of Gravity'',  \textit{ Phys.Rev.}, \textbf{D 81}, 124040  [arXiv:1003.5665].



\bibitem{A22}  Chakraborty S., T. Padmanabhan, 2014 ``Geometrical variables with direct thermodynamic significance in Lanczos-Lovelock gravity'', \textit{Phys. Rev.}, \textbf{D 90}, 084021  [arXiv:1408.4791] 




\bibitem{du1}  Davies P. C. W., 1975 ``Scalar production in Schwarzschild and Rindler metrics"  \textit{J. Phys.}, \textbf{A 8} 609.

\bibitem{du2}
Unruh W. G. , 1976 ``Notes on black-hole evaporation" \textit{Phys. Rev.} \textbf{D 14}, 870.



\bibitem{A1}
Bekenstein, J.D., {Black holes and entropy},
{\em Phys.~Rev. D} {\bf 1973}, {\em 7},
 2333.


 \bibitem{A4} 
Hawking, S., {Black Holes and Thermodynamics},
{\em Phys.~Rev. D} {\bf 1976}, {\em 13}, 191.
 

\bibitem{A8} 
Iyer, V., Wald, R.M., {Some properties of Noether charge and a proposal for dynamical black hole entropy}, {\em Phys.~Rev. D} {\bf 1994}, {\em 50}, 846.





\bibitem{dawood1} Kothawala D., ``The thermodynamic structure of Einstein tensor'', 2011	\textit{Phys.Rev.} \textbf{D 83}, 024026,   [arXiv:1010.2207]




\bibitem{tg}  Carroll S.M. and  Remmen G. N ,  `` What is the entropy in entropic gravity?'', [arXiv:1601.07558] 



\bibitem{sanvedtp}  Kolekar S., T. Padmanabhan, 2012 ``Action principle for the Fluid-Gravity correspondence and emergent gravity'', \textit{Phys.Rev.}, \textbf{D 85}, 024004  [arXiv:1109.5353]



\bibitem{h214}  Padmanabhan T., 2011 ``Entropy density of spacetime and the Navier-Stokes fluid dynamics of null surfaces'', \textit{Phys.Rev.}, \textbf{D 83}, 044048  [arXiv:1012.0119]


\bibitem{stefano}   Chirco G., C. Eling, S. Liberati,  2011 ``Reversible and Irreversible Spacetime Thermodynamics for General Brans-Dicke Theories'' \textit{Phys.Rev.} \textbf{ D83} 024032. [arXiv:1011.1405].


\bibitem{soloreview}  Solodukhin S. N., 2011 ``Entanglement entropy of black holes'', \textit{Living Rev. Relativity} \textbf{14},  8  [arXiv:1104.3712].


\bibitem{tpentangle}
 Padmanabhan T., 2010 ``Finite entanglement entropy from the zero-point-area of spacetime'', \textit{Phys.Rev.}, \textbf{D 82}, 124025  [arXiv:1007.5066]



\bibitem{mv}  Visser M., 2011 ``Conservative entropic forces'', \textit{JHEP} \textbf{10}  140 [arXiv:1108.5240]



\bibitem{tpentropy}  Padmanabhan T.,2015 ``Distribution function of the Atoms of Spacetime and the Nature of Gravity'', \textit{ Entropy} \textbf{17}, 7420-7452 [arXiv:1508.06286]




 \bibitem{key7}
 Padmanabhan T., 2010 \textit{Gravitation: Foundations and Frontiers}, (Cambridge University Press).




\bibitem{h191}  Padmanabhan T., 2008  ``Dark Energy and Gravity'' \textit{Gen.Rel.Grav.} \textbf{40} 529-564  [arXiv:0705.2533]



\bibitem{C7}
Padmanabhan, T., 2009 ``Dark Energy and its Implications for Gravity'', {\em Adv.~Sci.~Lett.}, {\bf 2}, 174--183.


\bibitem{D1} 
Kothawala, D., Padmanabhan, T., 2014 ``Grin of the Cheshire cat: Entropy density of spacetime as a relic from quantum gravity'', {\em Phys. Rev. D}, \textbf{90}, 124060. 
 
\bibitem{D4}
Kothawala, D., ``Minimal Length and Small Scale Structure of Spacetime'', 2013 {\em Phys. Rev. D}, \textbf{88}, 104029. 
 
\bibitem{D5}
 Stargen D. J., Kothawala, D., 2015, ``Small scale structure of spacetime: Van Vleck determinant and equi-geodesic surfaces'',  \textit{Phys. Rev.} \textbf{D 92}, 024046  [arXiv:1503.03793].

\bibitem{D6}
Kothawala, D., Padmanabhan, T, 2015 ``Entropy density of spacetime from the zero point length'', {\em Phys.~Lett. B}, \textbf{748}, 67--69. 




\bibitem{D8} 
Gray, A., ``The volume of a small geodesic ball of a Riemannian manifold'', 1974 {\em Mich. Math. J.}, \textbf{20},~329--344. 
 


\bibitem{D2a}  
 DeWitt B.S., 1964 ``Gravity: a Universal Regulator?'', \textit{Phys. Rev. Lett. } \textbf{13}, 114.

\bibitem{D2b} 
 Padmanabhan T., 1985  ``Planck length is the lower bound to all physical length scales'' \textit{Gen. Rel. Grav.} \textbf{17}, 215 

\bibitem{D2c} 
 Padmanabhan T., 1985  ``Physical significance of Planck length'' \textit{Ann. Phys.} \textbf{165}, 38 

\bibitem{D2d} 
 Padmanabhan T., 1997  ``Duality and zero-point length of spacetime'' \textit{Phys. Rev. Lett.}, \textbf{78}, 1854  [hep-th/9608182]

\bibitem{D2e}
 Garay L., 1998 ``Spacetime foam as a quantum thermal bath''\textit{ Phys. Rev. Lett.} \textbf{80}, 2508 [gr-qc/9801024]

\bibitem{D2f} 
 Garay L., 1995  ``Quantum gravity and minimum length'' \textit{Int. J. Mod. Phys. A}  \textbf{10}, 145.  



\bibitem{paperD} 
Padmanabhan, T., Chakraborty, S., Kothawala, D.,  2015, ``Renormalized spacetime is two-dimensional at the Planck scale', [arXiv:1507.05669].


\bibitem{D7} 
Chakraborty, S., Padmanabhan, T., [work in progress.]
 


\bibitem{z1}
Carlip,~S., Mosna,~R., Pitelli, J., 2011 ``Vacuum Fluctuations and the Small Scale Structure of Spacetime'',  \textit{{Phys. Rev. Lett.}}, \textit{107}, 021303. 

\bibitem{z2}
Ambjorn,~J., Jurkiewicz,~J., Loll,~R., 2005 ``Spectral Dimension of the Universe'', \textit{{Phys. Rev. Lett.}}, \textbf{95}, 171301.

\bibitem{z3}
Modesto,~L., 2009 ``Fractal Structure of Loop Quantum Gravity'', \textit{{Class. Quantum Grav.}}, \textbf{26}, 242002. 

\bibitem{z4}
Husain,~V., Seahra, S.S., Webster,~E.J, 2013 ``High energy modifications of blackbody radiation and dimensional reduction'', \textit{Phys. Rev. D}, \textbf{88}, 024014. 

\bibitem{nicolini}
Modesto, L and P.~Nicolini, 2010
`Spectral dimension of a quantum universe'',
  \textit{Phys.\ Rev.} {\bf D 81}, 104040 
  [arXiv:0912.0220]


\bibitem{kernela}
See e.g., D.J. Toms, 2007  ``The Schwinger Action Principle and Effective Action'',
(Cambridge University Press)

\bibitem{kernelb}
 Sriramkumar L., T. Padmanabhan, 2002 ``Probes of vacuum structure of quantum fields in classical backgrounds,'' \textit{Int.J.Mod.Phys.}, \textbf{D11}, 1-34  [gr-qc/9903054]



\bibitem{C8}
Padmanabhan, H., Padmanabhan, T., 2013 ``CosMIn: The solution to the cosmological constant problem'', \textit{Int.~J. Mod.~Phys. D}  {\bf 22}, 1342001.

\bibitem{C9} 
Padmanabhan, T., Padmanabhan, H., 2014  ``Cosmological constant from the emergent gravity perspective'', {\em Int.~J. Mod.~Phys. D}  {\bf 23}, 1430011. 




\end{thebibliography}
\end{document}